\documentstyle[sprocl,epsfig]{article}

\newcommand{\veps}{\mbox{\boldmath $\epsilon$ \unboldmath}}
\newcommand{\valf}{\mbox{\boldmath $\alpha$ \unboldmath}}
\newcommand{\vtau}{\mbox{\boldmath $\tau$ \unboldmath}}  
\newcommand{\vpi}{\mbox{\boldmath $\pi$ \unboldmath}}  
\def\Journal#1#2#3#4{{#1} {\bf #2}, #3 (#4)}

\def\NPB{{\em Nucl. Phys.} B}
\def\NPA{{\em Nucl. Phys.} A}
\def\PLB{{\em Phys. Lett.}  B}
\def\PRL{\em Phys. Rev. Lett.}
\def\PRD{{\em Phys. Rev.} D}

\def\PRC{{\em Phys. Rev.} C}

\def\be{\begin{equation}}
\def\ee{\end{equation}}
\def\bea{\begin{eqnarray}}
\def\eea{\end{eqnarray}}
\begin{document}

%To Prof Nick Karayiannis -- do read this:-
%If needed the word of Chapter~1, you can type in at the 
%\title{}. The words will be in caps and lowercase. 
%For chapter title can be in all caps or in caps and lowercase.
%It is up to the author to type for the case sensitive but 
%all articles must be in the same style. 
%But mostly for Review Volume are without this Chapter~1.
%Thank you
%Jessie   13/4/2000

\title{Study $\omega$ and $\phi$ photoproduction
 in the nucleon isotopic channels}

\author{Q. Zhao}

\address{Department of Physics, University of Surrey, Guildford, GU2 7XH, UK
\\E-mail: qiang.zhao@surrey.ac.uk} 

%\author{A. N. OTHER}

%\address{Department of Physics, Theoretical Physics, 1 Keble Road,\\
%Oxford OX1 3NP, England\\E-mail: other@tp.ox.uk}


\maketitle\abstracts{ 
We present results for the photoproduction of $\omega$ and $\phi$
meson in the nucleon isotopic channles. 
A recently developed quark model with an effective Lagrangian
is employed to account for the non-diffractive {\it s}- and {\it u}-channel
processes; the diffractive feature arising from the {\it natural}
parity exchange 
is accounted for by the {\it t}-channel pomeron 
exchange, while the {\it unnatural} parity exchange
is accounted for by the {\it t}-channel pion exchange. 
In the $\omega$ production, the isotopic effects
could provide more information concerning the 
search of ``missing resonances", while 
in the $\phi$ production, the isotopic effects
could highlight non-diffractive resonance excitation mechanisms 
at large angles. }

\section{Introduction}

The availabilities of high intensity electron and photon beams
at JLab (CLAS), ELSA (SAPHIR), MAMI, ESRF (GRAAL), and 
SPring-8 give accesses to excite nucleons
with the clean electromagnetic probes, thus revive
the interest in the search of ``missing resonances"~\cite{NRCQM-1}
in meson photo- and electroproduction. 
Vector meson photoproduction near threshold attracts 
attentions~\cite{zhao-98,zhao-prc-01,oh-01,oh,titov} 
since those ``missing resonances" could 
have stronger couplings to this channel such that 
their existing signals might be derived. 

Taking into account large degrees of freedom in the resonance excitations,
experimental efforts for such a purpose is by no means 
trivial. 
Historically, experimental data for photoproduction
of vector meson in isotopic channels are very sparse. 
In contrast with the reaction $\gamma +p\to \omega (\rho^0)+ p$,
there are only few dottes available for
$\gamma + n\to \rho^- + p$ and $\gamma + p\to \rho^+ + n$, while 
experiment on $\gamma + n\to\omega(\phi) + n$ is not available.
The main feature in the latter reactions is that
the electric transition vanishes in the nucleon pole terms. 
Nevertheless, due to the change of the isospin degrees of freedom, 
interferences between different resonances and processes
will change significantly. Nevertheless, in the neutron target
reaction, resonances of quark model
representation $[{\bf 70}, \ ^{\bf 4} {\bf 8}]$
are no longer suppressed by the Moorhouse selection rule~\cite{moorhouse}, 
thus will add more ingredients into the reaction. 
Briefly, apart from various polarization observables
which could provide rich information about the reaction mechanism, 
coherent study of isotopic reactions could also highlight
signals which cannot be seen easily in a single channel.

Our motivation of studying 
the isotopic production of the $\phi$ meson
is rather different from the $\omega$.
Our attention here is paid to the 
effects from the non-diffractive {\it s}- and {\it u}-channel
$\phi$ production in polarization observables.
The study of isotopic reaction provides us with 
insights into the non-diffractive processes at large angles (large
momentum transfer with $W\approx 2\sim 4$ GeV).
Since the $\phi$ meson has higher threshold ($E_\gamma\approx 1.57$ GeV), 
where a large number of resonances can be excited,
the non-diffractive resonance effects 
could exhibit more collectively rather than 
exclusively from individual resonances. 
Another relevant interest in $\phi$ production is
to detect strangeness component in nucleons, which could be another
important non-diffractive source in the 
reaction~\cite{henley92,titov97-prl,titov98}.
Although our study does not take into account such a mechanism,
we shall see that our results provide some 
supplementary information for such an effort.

In this proceeding, a quark model
approach to vector meson photoproduction
~\cite{zhao-98,zhao-prc-01,zhao-phi-99,zhao-plb-01}
is applied to the photoproduction of $\omega$ and $\phi$ meson
off the proton and neutron. 
Our purpose is to provide
a framework on which a systematic study of 
resonance excitations becomes possible.

\section{The model}

Our model
consists of three processes: (i) 
vector meson production through an effective Lagrangian;
(ii) {\it t}-channel pomeron exchange ($\cal{P}$) for $\omega$,
$\rho^0$ and $\phi$ production~\cite{donnachie}; 
(iii) {\it t}-channel
light meson exchange. 
Namely, in the $\omega$ and $\phi$ meson photoproduction, 
the $\pi^0$ exchange ($\pi$)
is taken into account.~\footnote{In the $\rho^0$ production, 
the $\sigma$ meson exchange is included.} 

The effective Lagrangian for the $V$-$qq$ coupling is:
\begin{equation} \label{lagrangian}  
L_{eff}= \overline{\psi}(a\gamma_\mu +  
\frac{ib\sigma_{\mu\nu}q^\nu}{2m_q}) V^\mu \psi,
\end{equation}  
where $\psi$ ($\overline{\psi}$) denotes 
the light quark (anti-quark) field in a baryon system; 
$V^\mu$ represents vector meson field 
($\omega$, $\rho$, $K^*$ and $\phi$).   
In this model, the 3-quark baryon system is described by the NRCQM in the 
$SU(6)\otimes O(3)$ symmetry limit, while
the vector meson is treated as an elementary point-like particle 
which couples to the constituent quark through the effective interaction.
Parameter $a$ and $b$ denote the coupling strengths
and are determined by experimental data.

\begin{figure}[t]
%\rule{5cm}{0.2mm}\hfill\rule{5cm}{0.2mm}
%\vskip 2.5cm
%\rule{5cm}{0.2mm}\hfill\rule{5cm}{0.2mm}
\epsfig{figure=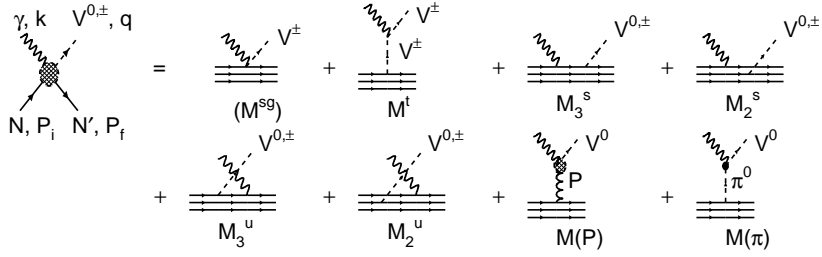,height=2.4in}
\caption{Tree level transition diagrams for the reaction.  \label{fig-1}}
\end{figure}

The tree level transitions from the effective Lagrangian 
can be labelled by the Mandelstam variable, {\it s}, {\it u}, and {\it t}:
\begin{equation}\label{trans-ampli}  
M_{fi}=M^s_{fi}+M^u_{fi}+M^t_{fi} \ .
\end{equation}  
Given the electromagnetic coupling 
$H_e=-\overline{\psi}\gamma_\mu e_q A^\mu\psi$, 
and the effective strong coupling
$H_m=-\overline{\psi}(a\gamma_\mu +\frac{ib\sigma_{\mu\nu}  
q^\nu}{2m_q}) V^\mu \psi$, 
the {\it s}- and {\it u}-channel with resonance excitations
can be expressed as,
\begin{eqnarray}  
M^{s+u}_{fi}&=&\sum_{j}\langle N_f|H_m|N_j\rangle\langle   
N_j|\frac{1}{E_i+\omega_\gamma-E_j}H_e|N_i\rangle\nonumber\\  
&&+\sum_{j} \langle N_f|H_e\frac{1}{E_i-\omega_m-E_j}  
|N_j\rangle\langle N_j|H_m|N_i\rangle, 
\end{eqnarray}
where $\omega_\gamma$ and $\omega_m$ represent the photon and meson
energy, respectively. In this expression, 
all the intermediate states $|N_j\rangle$ are included.
A simple transformation for the electromagnetic 
interaction leads to,
\begin{eqnarray}  
M^{s+u}_{fi}&=&i\langle N_f|[g_e, \ H_m]|N_i\rangle\nonumber\\
&&+i\omega_\gamma\sum_{j}\langle N_f|H_m|N_j\rangle\langle   
N_j|\frac{1}{E_i+\omega_\gamma-E_j}h_e|N_i\rangle\nonumber\\  
&&+i\omega_\gamma\sum_{j} \langle N_f|h_e\frac{1}{E_i-\omega_m-E_j}  
|N_j\rangle\langle N_j|H_m|N_i\rangle, 
\label{s+u}  
\end{eqnarray} 
with
$
h_e=\sum_{l}e_l {\bf r}_l\cdot\veps_\gamma(1-\valf\cdot  
{\hat{\bf k}})e^{i{\bf k\cdot r}_l}
$, and
$
g_e=\sum_{l}e_l{\bf r}_l\cdot\veps_\gamma e^{i{\bf k\cdot r}_l}
$,
where ${\bf{\hat k}}\equiv {\bf k}/\omega_\gamma$ is 
the unit vector along the photon three-momentum ${\bf k}$.
We identify the term $i\langle N_f|[g_e, \ H_m]|N_i\rangle$ 
as a seagull term $M^{sg}_{fi}$, and re-define the second and third
term as the {\it s}- and {\it u}-channel, respectively. 
The tree level diagrams are illustrated in Fig.~\ref{fig-1}, 
where a bracket on the seagull term gives a caution about
its origin.

In the NRCQM, those low-lying states ($n\leq 2$)
have been successfully related to the resonances which can be 
taken into account explicitly in the formula. For those higher 
excited states, they can be treated degenerate in the main quantum 
number $n$ in the harmonic oscillator basis. Detailed description 
of this approach can be found in Ref.~\cite{zhao-98,zhao-phi-99,zhao-prc-01}. 
The $\omega$ and $\phi$ production are significantly simplified
because of isospin conservation. Namely, only isospin 1/2 intermediate 
resonances will contribute in the reaction. 
Another advantage for these two reactions is that both particles
are charge-neutral. This feature leads to the vanishing 
of the {\it t}-channel vector meson exchange and 
the seagull term.
In the end, only the {\it s}- and {\it u}-channel (S+U)
from Eq.~\ref{s+u}
will be the contribution source from the effective Lagrangian.
In addition, the Moorhouse selection rule~\cite{moorhouse} 
further simplifies the calculations for the proton target reaction. 
Due to this selection rule, resonances belonging to
representation $[{\bf 70}, ^{\bf 4}{\bf 8}]$ would vanish
in the $\gamma p\to N^*$ transitions.

%%%%%%%% FIG 1 
%\vspace*{-1.5cm}

Apart from the S+U from the effective Lagrangian,
our model includes the pomeron exchange ($\cal{P}$) to account for 
the diffractive phenomenon in vector meson photoproduction.
In this model~\cite{donnachie}, the pomeron  
mediates the long range interaction between 
two confined quarks, and behaves rather like a $C=+1$ isoscalar photon. 
We summarize the vertices as follows:

(i) pomeron-nucleon coupling: 
\begin{equation}
F_{\mu}(t)= 3\beta_0\gamma_{\mu}f(t), \  
f(t)= \frac{(4M^2_N-2.8t)}{(4M^2_N-t)(1-t/0.7)^2} \ ,
\end{equation}
where $\beta_0=1.27$ GeV$^{-1}$ 
is the coupling of the pomeron to one light constituent quark; 
$f(t)$ is the isoscalar nucleon electromagnetic form factor 
with four-momentum transfer $t$;
the factor 3 comes from the ``quark-counting rule".

(ii) Quark-$\phi$-meson coupling: 
\begin{equation}
V_\nu(p-\frac 12 q, p+\frac 12 q)=f_V M_\phi\gamma_\nu \ ,
\end{equation}
where $f_V$ is the radiative decay constant 
of the vector meson, and
determined by 
$\Gamma_{V\to e^+e^-}=8\pi \alpha^2_e e^2_Q f^2_V/3M_V$.
A form factor $\mu^2_0/(\mu^2_0+p^2)$ is adopted 
for the pomeron-off-shell-quark vertex,
where $\mu_0=1.2$ GeV is the cut-off energy, 
and $p$ is the four-momentum of the quark.
The pomeron trajectory is
$\alpha(t)=1+\epsilon+\alpha^\prime t $, 
with $\alpha^\prime=0.25$ GeV$^{-2}$.

The $\pi^0$ exchange is introduced with
the following Lagrangians for the $\pi NN$ 
and $\phi\pi\gamma$ coupling:
\begin{eqnarray}\label{3}
L_{\pi NN}=-i g_{\pi NN}\overline\psi \gamma_5(\vtau\cdot\vpi)\psi \ .
\end{eqnarray}
and 
\begin{eqnarray}\label{4}
L_{V \pi^0 \gamma}=e_N\frac{ g_{V\pi\gamma} }{M_V}
\epsilon_{\alpha\beta\gamma\delta}\partial^\alpha A^\beta
\partial^\gamma V^\delta\pi^0 \ .
\end{eqnarray}
where the commonly used couplings, 
${g^2_{\pi NN}}/{4\pi}=$ 14,
$g^2_{\omega\pi\gamma}=3.315$ and $g^2_{\phi\pi\gamma}=0.143$, are adopted.

An exponential factor 
$e^{-({\bf q}-{\bf k})^2/6\alpha^2_\pi}$ from the nucleon wavefunctions
plays a role as a form 
factor for the $\pi NN$ and $\phi\pi\gamma$ vertices, 
where $\alpha_\pi=300$ MeV is adopted.
This factor comes out naturally in the harmonic oscillator basis
where the nucleon is treated as a 3-quark system.

\section{Results and discussions}

Application of this approach to the $\omega$ meson photoproduction
shows a great promise. A coherent study of the 
$\rho$ meson photoproduction~\cite{zhao-98} also highlights the 
approximately satisfied isospin symmetry between 
the $\omega$ and $\rho$ meson. 

In $\gamma p\to \omega p$, the old measurements~\cite{ballam73}
of the parity asymmetries at $E_\gamma=$ 2.8, 4.7 and 9.3 GeV 
provides very strong constraints on the {\it natural} ($\cal{P}$)
and {\it unnatural}-parity exchange ($\pi$).
It was shown that at $E_\gamma=2.8$ GeV, the parity asymmetry
has a negative value which suggests the still dominant
contribution from the pion exchange. At 4.7 GeV, the {\it natural}-parity
starts to dominate over the cross section, and at 9.3 GeV, 
the parity asymmetry is completely dominated by the pomeron exchanges. 
Such an energy evolution of the parity asymmetry
above the resonance region constrains the ``background" terms, 
which then can be extrapolated down to the resonance region. 
In Ref.\cite{zhao-prc-01,zhao-nstar}, 
we show that such a scheme is
consistent with the experimental data 
for the parity asymmetries~\cite{ballam73}, 
and indeed leads to a reasonable estimation of these ``background"
terms.

\subsection{$\omega$ meson photoproduction}

Parameters at resonance region for the $\omega$ photoproduction are constrained
approximately by the SAPHIR data~\cite{klein}.
It shows that $a=-2.72$ and $b^\prime\equiv b-a=-3.42$ 
give an overall fit of available data.

\begin{figure}[t]
\begin{center}
\epsfig{file=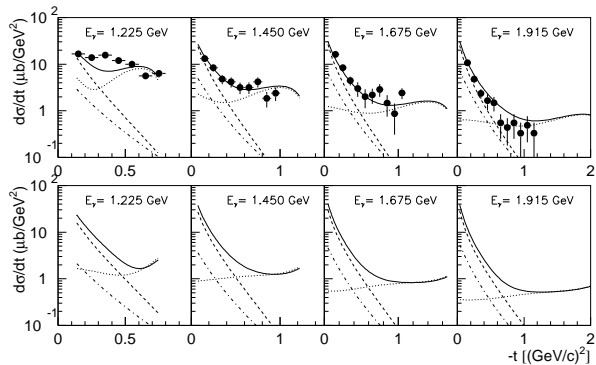,height=2.2in}
\end{center}
\caption{ Differential cross section for $\gamma p \to \omega p$ (upper column)
and $\gamma n \to \phi n$ (lower column). 
The dashed, dot-dashed, and dotted curves
denote exclusive results for process, S+U, $\cal{P}$ and $\pi$, respectively;
the solid curves 
represent full model calculation.
Data are from Ref.~\protect\cite{klein}. \protect\label{fig-2}}
\end{figure}

In Fig.~\ref{fig-2}, differential 
cross sections for the isotopic channels are presented at
four energy bins, $E_\gamma=$1.225,  
1.450, 1.675 and 1.915 GeV, and compared 
with the SAPHIR data~\cite{klein}. 
It shows that near threshold the 
$\pi^0$ exchange
($\pi$)
dominates over the other two processes at small angles, 
while $\cal{P}$ will compete with $\pi$
with the increasing energies. 
The exclusive $\cal{P}$ and $\pi$ in the proton and 
neutron reaction are changed slightly due to the small 
difference between the proton and neutron mass.
The S+U contributions 
generally dominate at large angles. 
Compare these two isotopic reactions, 
we see that the main impact of the change of the isospin space 
comes from S+U. 

%%%%%%%% FIG 2

\begin{figure}[t]
%\begin{center}
\epsfig{file=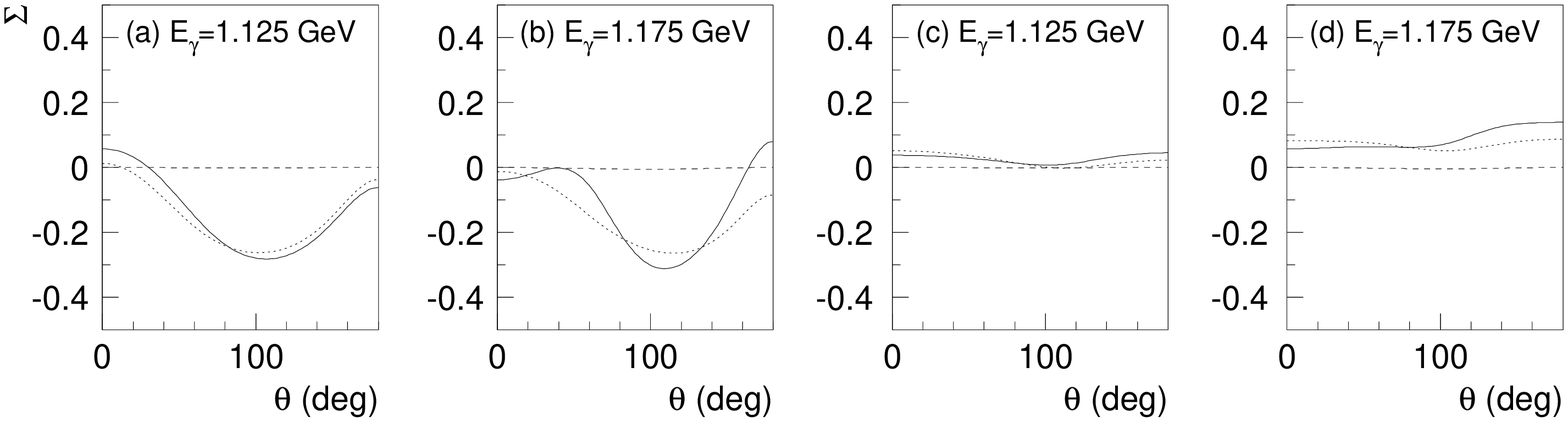,height=1.5in}
%\end{center}
\caption{ Polarized beam symmetry
for the proton [(a)-(b)] and neutron reactions [(c)-(d)] 
in $\omega$ production. 
The solid, dashed, and dotted curves
denote the full model calculation, $\cal{P}+\pi$, 
the full model
excluding the $P_{13}(1720)$. \protect\label{fig-3} }
\end{figure}

Interesting interfering effects 
among these processes could be seen in polarization observables.
In Fig.~\ref{fig-3}, we present predictions for 
$\Sigma\equiv (2\rho^1_{11}+\rho^1_{00})/(2\rho^0_{11}+\rho^0_{00})$,
which has been measured by GRAAL~\cite{graal}. Here, 
$\rho^1$ and $\rho^0$ are density matrix elements in the helicity 
space~\cite{schilling}.
One important feature of $\Sigma$ is that 
large asymmetries (deviation from 0) cannot be produced by the 
exclusive $\cal{P}$, $\pi$ or $\cal{P}+\pi$.
As shown by the solid curves, large asymmetries are produced by
the interferences between the $\cal{P}+\pi$ and 
S+U processes. Interest arises from the comparison of the isotopic reactions. 
It shows that the S+U produces flattened positive 
asymmetries in $\gamma n\to\omega n$ near threshold. 
Specifically, we show the effects of $P_{13}(1720)$ 
in the $\Sigma$, which suggest that isotopic reactions
could have outlined more information.

%%%%%%%% FIG 3

Another interesting observable with polarized photon beam is,
$\Sigma_A\equiv (\rho^1_{11}+\rho^1_{1-1})/(\rho^0_{11}+\rho^0_{1-1})$.
As found in Ref.~\cite{zhao-prc-01}, this observable 
is more sensitive to small contributions from individual resonances.
Due to lack of space, we shall only show the calculation of $\Sigma_A$ 
for the $\phi$ production in the next Subsection.

\subsection{$\phi$ meson photoproduction}

In the $\phi$ meson production, 
the parameter $a=0.241\pm 0.105$ and $b^\prime\equiv b-a=0.458\pm 0.091$ 
for the effective $\phi$-$qq$ coupling
are determined by the fit of old data from Ref.~\cite{besch74}
at $E_\gamma=2.0$ GeV
and the recent data from CLAS~\cite{clas-phi-00} at 3.6 GeV. 
These values are much smaller than parameters derived 
for the $\omega$ and $\rho$ production.
Qualitatively, the $\phi$-$qq$ couplings could be suppressed by
the OZI rule. Meanwhile, in the SU(3) symmetry limit,
$a_\phi/a_\omega= g_{\phi NN}/g_{\omega NN}$ can be derived, 
which is consistent with the ratio for 
an ideal $\omega$-$\phi$ mixing~\cite{PDG2000},
$g_{\phi NN}/g_{\omega NN}=-\tan 3.7^\circ$.

%%%%%%%% FIG 4
\begin{figure}[t]
%\begin{center}
\begin{minipage}[t]{54mm}
\epsfig{file=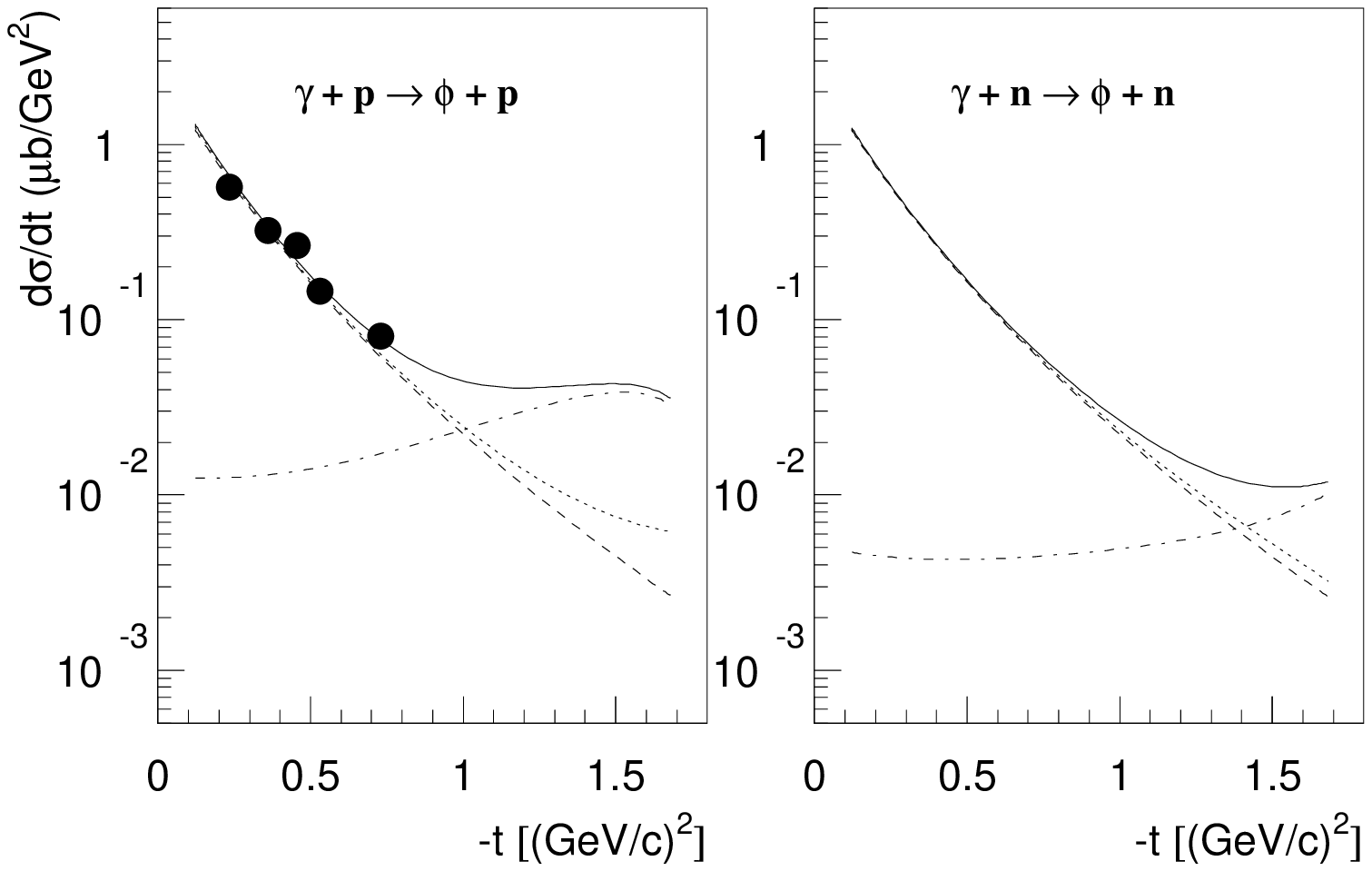,height=1.6in}
\caption{ Differential cross section for $\phi$ production
at $E_\gamma=2.0$ GeV. 
The dot-dashed, dashed, and solid curves
denote the S+U, $\cal{P}+\pi$, and full model 
calculations, respectively, while the dotted curve 
represents full model calculation
excluding the {\it u}-channel contribution.
Data come from Ref.~\protect\cite{besch74}. \protect\label{fig-4}}
\end{minipage}
%\end{center}
%
%
\hspace{\fill}
%\end{figure}
%
%%%%%%%% FIG 5
%\begin{figure}[t]
%\begin{center}
\begin{minipage}[t]{54mm}
\epsfig{file=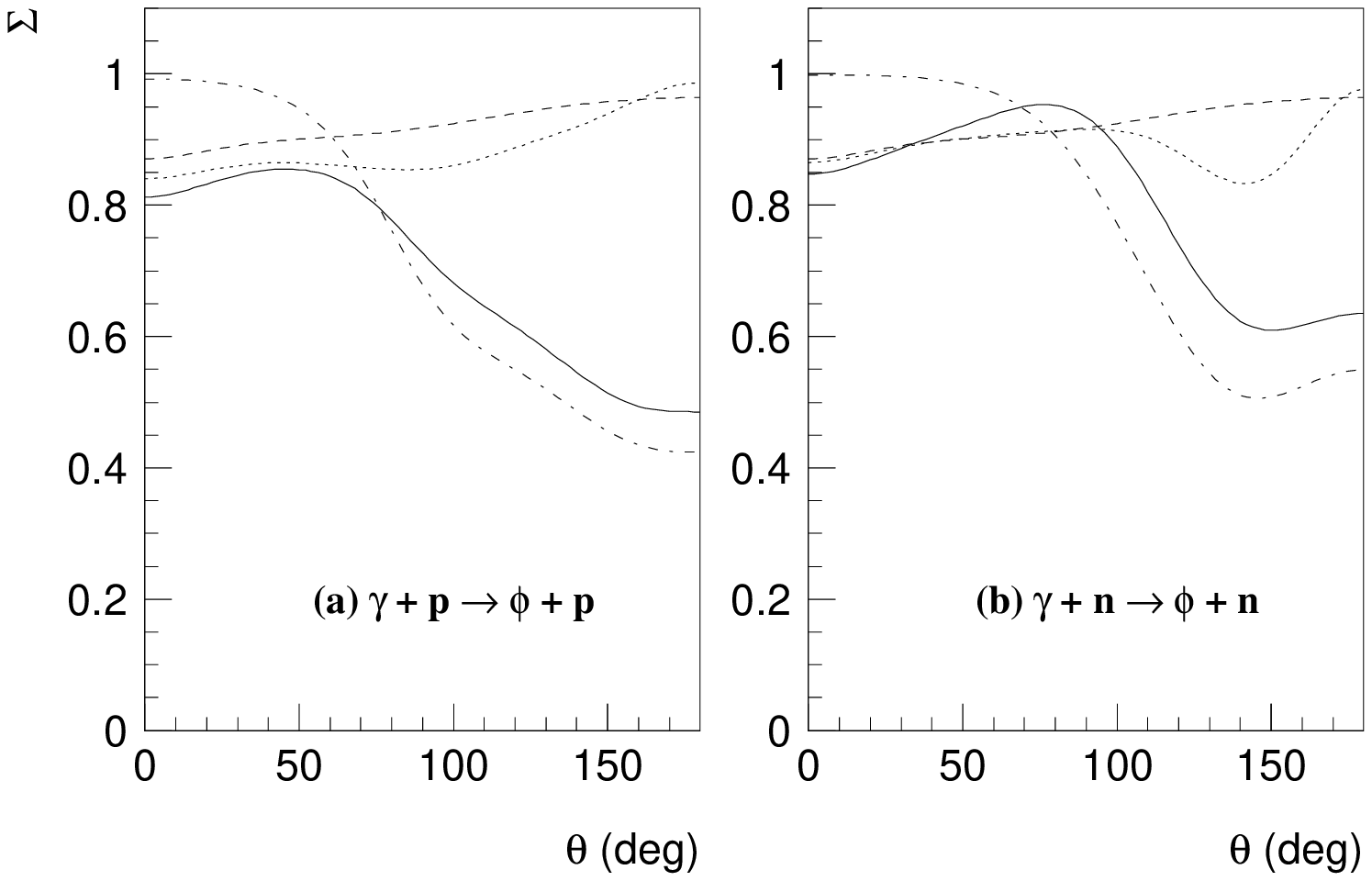,height=1.6in}
\caption{ Polarized beam symmetry $\Sigma_A$
for the proton and neutron reactions at $E_\gamma=2.0$ GeV. 
Notations are the same as Fig.~\protect\ref{fig-4}. \protect\label{fig-5}}
\end{minipage}
%\end{center}
\end{figure}
Using the same parameters derived in $\gamma p \to \phi p$, 
the cross sections for both $\gamma p \to \phi p$ and $\gamma n \to \phi n$ are 
calculated at $E_\gamma=2.0$ GeV (Fig.~\ref{fig-4}). 
Significant cross section difference occurs at large angles, where
$d\sigma/d\Omega$ for $\gamma n \to \phi n$ is much smaller
than for $\gamma p \to \phi p$. 
A common feature arising from these two reactions 
is a relatively stronger backward peak 
from the ``background" {\it u}-channel nucleon pole term.
Similar feature is also found by Laget~\cite{laget00}.

In Fig.~\ref{fig-5}, 
the isotopic effects 
of these two reactions 
are shown for the polarized beam asymmetry
$\Sigma_A\equiv (\rho^1_{11}+\rho^1_{1-1})/(\rho^0_{11}+\rho^0_{1-1})
=(\sigma_\parallel-\sigma_\perp)/(\sigma_\parallel+\sigma_\perp)$ 
at $E_\gamma=2.0$ GeV,
where $\sigma_\parallel$ and $\sigma_\perp$
denote the cross sections for $\phi\to K^+K^-$ 
when the decay plane is parallel or perpendicular to 
the photon polarization vector.

The dashed curves represent results for 
$\cal{P}+\pi$, which 
deviate
from $+1$ (for pure {\it natural}-parity exchange) 
due to the presence of 
the {\it unnatural} parity pion exchange.
With the S+U,
the full model calculation suggests that the large angle
asymmetry is strongly influenced by the S+U processes, while 
the forward angles are not sensitive to them.
Interferences between 
the $\cal{P}$ and S+U  
can be seen by excluding the $\pi$ (see the dot-dashed curves). 
It shows that asymmetries produced by the S+U at forward angles 
are negligible.
Since the pion exchange becomes 
very small at large angles, we conclude that
the large angle asymmetry is determined by the 
S+U and reflects
the isotopic effects.
The role played by the {\it s}-channel resonances
in the two reactions are presented by excluding
the {\it u}-channel contributions.
As shown by the dotted curves, the interferences from
the {\it s}-channel resonances are much weaker  
than that from the {\it u}-channel, although they are still 
an important non-diffractive source at large angles.

This feature, which 
might make it difficult to filter out signals for individual 
{\it s}-channel resonances in the $\phi$ photoproductions
 might suggest that the forward angle could be an ideal region 
for the study of 
other non-diffractive sources.~\footnote{ For example, it was suggested
by Ref.~\protect\cite{henley92,titov97-prl}, 
the strangeness component in nucleons could produce 
large asymmetries in beam-target double polarization
observable at forward angles, while we find that S+U process
only make sense at large angles.~\protect\cite{zhao-phi-99,zhao-plb-01} }

It should be noted that no isotopic effects 
can be seen in $\Sigma$ and $\Sigma_A$ if only $\cal{P}+\pi$
contribute to the cross section. 
This is because the transition amplitude 
of $\cal{P}$ is purely imaginary, while
that of pion exchange is purely real. 
In $\Sigma$ and $\Sigma_A$, the sign arising from the 
$g_{\pi NN}$ will disappear, which is why
the dashed curves in Fig.~\ref{fig-5} are (almost) the same.
It is worth noting that our results
for the $\Sigma_A$ are 
quite similar to 
findings of Ref.~\cite{titov99} at small angles,
but very different at large angles. 
In Ref.~\cite{titov99} only the 
nucleon pole terms for the {\it s}- and 
{\it u}-channel processes were included.

\section{Conclusions and perspectives}

We show that the effective Lagrangian approach in the quark model basis
provides an ideal framework for the study of resonance excitations
in vector meson photoproduction. 
The great advantage is that only a limited number of parameters
will be introduced. Meanwhile, it permits a systematic and coherent 
study of different isospin channels, which could be helpful 
in highlighting model-independent features in the reaction mechanisms.
In the SU(6)$\otimes$O(3) framework, 
photoproduction of isospin 1 ($\rho^{0,\pm 1}$)
and isospin 1/2 ($K^{*\pm}$, $K^{*0}$ and $\overline{K}^{*0}$) 
can be also studied~\cite{zhao-98,zhao-kstar-01}.

Conerning the search of 
signals for $s\overline{s}$ component 
in the nucleon, the forward angle kinematics 
might be selective if the findings of Refs.~\cite{titov97-prl,titov98}
are taken into account, 
since effects from the S+U
are negligible. 
Certainly, since a possible strangeness content has not been 
explicitly included in this model, the effective $\phi$-$qq$ coupling 
cannot distinguish between an OZI evading $\phi NN$ coupling
and a strangeness component in the nucleon. 
In future study, a more complex approach including 
the possible strangeness component in the nucleon should be explored.

Concerning the efforts of searching for ``missing resonances" 
(in $\omega$ photoproduction)~\cite{cole} and 
probing strangeness components (in $\phi$)~\cite{nakano,tedeschi}, 
it shows that the isotopic channels might 
be useful for disentangling various processes involved in the reaction
mechanism. 
Experiments using the neutron target 
could provide more evidence to pin down model-independent aspects 
for any modelling~\cite{didelez}.

\section*{Acknowledgments}
I thank J.-P. Didelez, M. Guidal, 
B. Saghai, and J.S. Al-Khalili
for collaborations on part of the work. 
Fruitful discussions with E. Hourany concerning the GRAAL 
experiment, T. Nakano concerning the SPring-8, 
and P.L. Cole and D.J. Tedeschi concerning the CLAS 
experiments, are acknowledged.

\end{document}